\pdfoutput=1

\documentclass[aps,prl,amsmath,amssymb,floatfix,twocolumn,amsmath,superscriptaddress,twocolumn,nofootinbib,tighten,letterpaper]{revtex4-2}

\usepackage[utf8]{inputenc}
\usepackage{multirow}

\usepackage{enumerate}
\usepackage{amsfonts}
\usepackage{amsmath}
\usepackage{amssymb}
\usepackage{color}
\usepackage{bm}
\usepackage{graphicx}
\usepackage[caption=false]{subfig}

\usepackage{tabularray}
\definecolor{RowColor}{rgb}{0.88,1,0.9}

\let\phi=\varphi
\let\epsilon=\varepsilon

\renewcommand{\d}[1]{\mathrm{d}{#1}}

\definecolor{DarkRed}{rgb}{0.80,0,0}
\definecolor{DarkGray}{rgb}{0.7,0.7,0.7}

\newcommand{\prlsection}[1]{\textit{#1}.\kern0.05em---\kern0.05em\ignorespaces}
\begin{document}
\title{Yukawa-Lorentz Symmetry in Non-Hermitian Dirac Materials}

\author{Vladimir Juri\v{c}i\'c}~\thanks{Corresponding Author: vladimir.juricic@su.se}
\affiliation{Departamento de F\'isica, Universidad T\'ecnica Federico Santa Mar\'ia, Casilla 110, Valpara\'iso, Chile}
\affiliation{Nordita, KTH Royal Institute of Technology and Stockholm University,
Hannes Alfvéns väg 12, SE-106 91 Stockholm, Sweden}

\author{Bitan Roy}~\thanks{Corresponding Author: bitan.roy@lehigh.edu}
\affiliation{Department of Physics, Lehigh University, Bethlehem, Pennsylvania, 18015, USA}

\maketitle


{\bf Lorentz spacetime symmetry represents a unifying feature of the fundamental forces, typically manifest at sufficiently high energies, while in quantum materials it emerges in the deep low-energy regime.  However, its fate in quantum materials coupled to an environment thus far remained unexplored.  We here introduce  a general framework of constructing symmetry-protected Lorentz invariant non-Hermitian (NH) Dirac semimetals (DSMs), realized by invoking masslike anti-Hermitian Dirac operators to its Hermitian counterpart. Such NH DSMs feature purely real or imaginary isotropic linear band dispersion, yielding a vanishing density of states. Dynamic mass orderings in NH DSMs thus take place for strong Hubbardlike local interactions through a quantum phase transition, hosting a non-Fermi liquid, beyond which the system becomes an insulator. We show that depending on the internal Clifford algebra between the NH Dirac operator and candidate mass order-parameter, the resulting quantum-critical fluid either remains coupled with the environment or recovers full Hermiticity by decoupling from the bath, while always enjoying an emergent Yukawa-Lorentz symmetry in terms of a unique terminal velocity. We showcase the competition between such mass orderings, their hallmarks on quasiparticle spectra in the ordered phases, and the relevance of our findings for correlated designer NH Dirac materials.}



From classical and quantum electrodynamics involving photons to quantum chromodynamics encompassing gluons, field theoretic formulations of fundamental forces rest on the unifying bedrock of the Lorentz symmetry~\cite{Jackson1999, Peskin2019}, typically realized at sufficiently high energies~\cite{Nielsen1978, Chadha1983, Kostelecky2011, Horava2011, Anber2011, Bednik2013}. On the other hand, in Dirac crystals, realized in a number of quantum materials~\cite{Castroneto2009, Balatsky2014, Armitage2018}, although such a symmetry may not necessarily be present at the microscopic level, it rises as an emergent phenomenon in the deep infrared regime through boson mediated inter-particle interactions~\cite{Gonzalez1994, SSLee2007, Isobe2012, Roy2016, Roy2018, Roy2020}. The bosonic degrees of freedom can be vectorlike particles such as helicity-1 photons which interact with Dirac fermions and charged Cooper pairs or spinless scalar order-parameter fluctuations at the brink of dynamic mass generation for Dirac quasiparticles via spontaneous symmetry breaking. Irrespective of these microscopic scenarios, the emergent Lorentz symmetry always manifests a unique velocity in the medium, generically tagged as the `speed of light', not necessarily $c$. Necessity and elegance of this commonly occurring space-time symmetry often allow us to take it for granted. However, its fate when quantum materials interact with the environment thus far remains an untouched territory, which we here set out to explore.

In the Hamiltonian language, system-to-environment couplings can be modeled by non-Hermitian (NH) operators. But, a one-to-one correspondence between them is still missing. Despite this limitation, we showcase a possible emergent Lorentz symmetry in nonspatial symmetry protected NH Dirac semimetals (DSMs), captured by Lorentz invariant NH Dirac operators, possessing purely real or imaginary eigenvalue spectra. When such an NH DSM arrives at the shore of dynamic mass generation triggered by Hubbardlike finite-range Coulomb repulsion, mediating boson-fermion Yukawa interactions, the resulting strongly coupled incoherent non-Fermi liquid  always features  a unique terminal velocity in the deep infrared regime. Depending on the pattern of the incipient Dirac insulation, the system achieves the Yukawa-Lorentz symmetry either maintaining its coupling with the environment or by decoupling itself from the bath. These outcomes possibly suggest a generic Lorentz symmetry in NH Dirac materials, despite the variety of its interactions with the environment.

\begin{figure}[t!]
\includegraphics[width=1.00\linewidth]{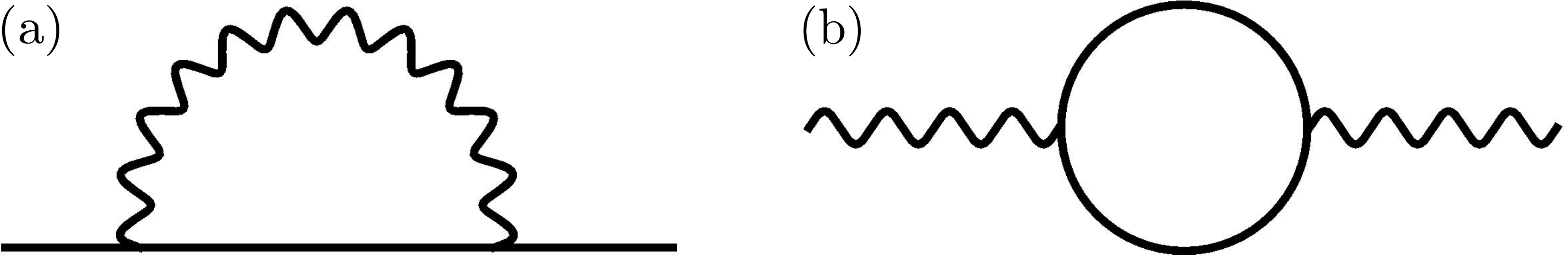}
\caption{{\bf Self-energy diagrams.} (a) Dirac fermion (solid lines) self-energy diagram. (b) Self-energy diagram for the  bosonic order parameter field (wavy lines). The vertex corresponds to the Yukawa coupling in Eq.~\eqref{eq:yukawa}.
}~\label{fig:Self-energy}
\end{figure}

\begin{figure}[t!]
\includegraphics[width=0.95\linewidth]{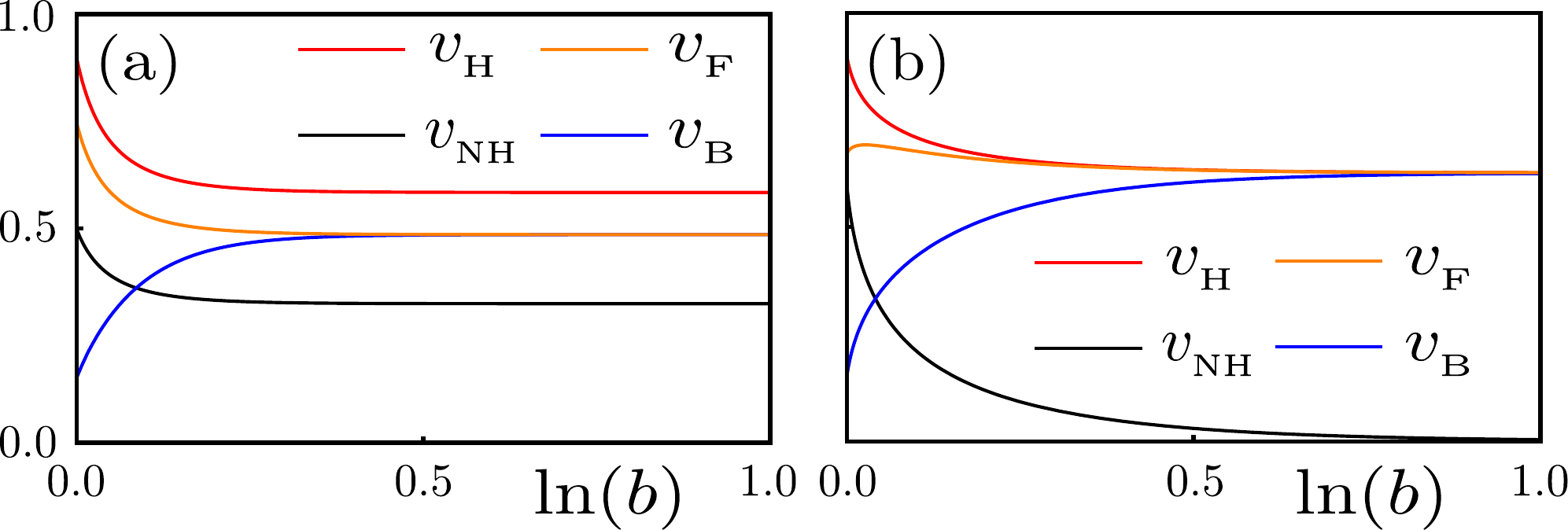}
\caption{{\bf Renormalization group flow of the velocities in the critical plane of the Gross-Neveu-Yukawa theory.} The flow for (a) commuting class mass (CCM) [Eqs.~\eqref{eq:velocity-fermi}-\eqref{eq:flow-comm-2}] and (b) anticommuting class mass (ACM) [Eqs.~\eqref{eq:velocity-fermi}, \eqref{eq:flow-anticomm-1} and \eqref{eq:flow-anticomm-2}] ordering. In (a), we set the bare values of the Hermitian , non-Hermitian and bosonic velocity parameters, respectively, $v_{_{\rm H}}^0=0.90$, $v_{_{\rm NH}}^0=0.50$, $v_{_{\rm B}}^0=0.15$,  while in (b) we use $v_{_{\rm H}}^0=0.90$, $v_{_{\rm NH}}^0=0.60$, $v_{_{\rm B}}^0=0.15$. Throughout we set the number of four-component Dirac fermion species $N_f=1$, the number of the bosonic order-parameter components $n=1$ and the Yukawa coupling constant $g^2=1$.}~\label{fig:TerminalVelocity}
\end{figure}

\begin{figure}[t!]
\includegraphics[width=0.95\linewidth]{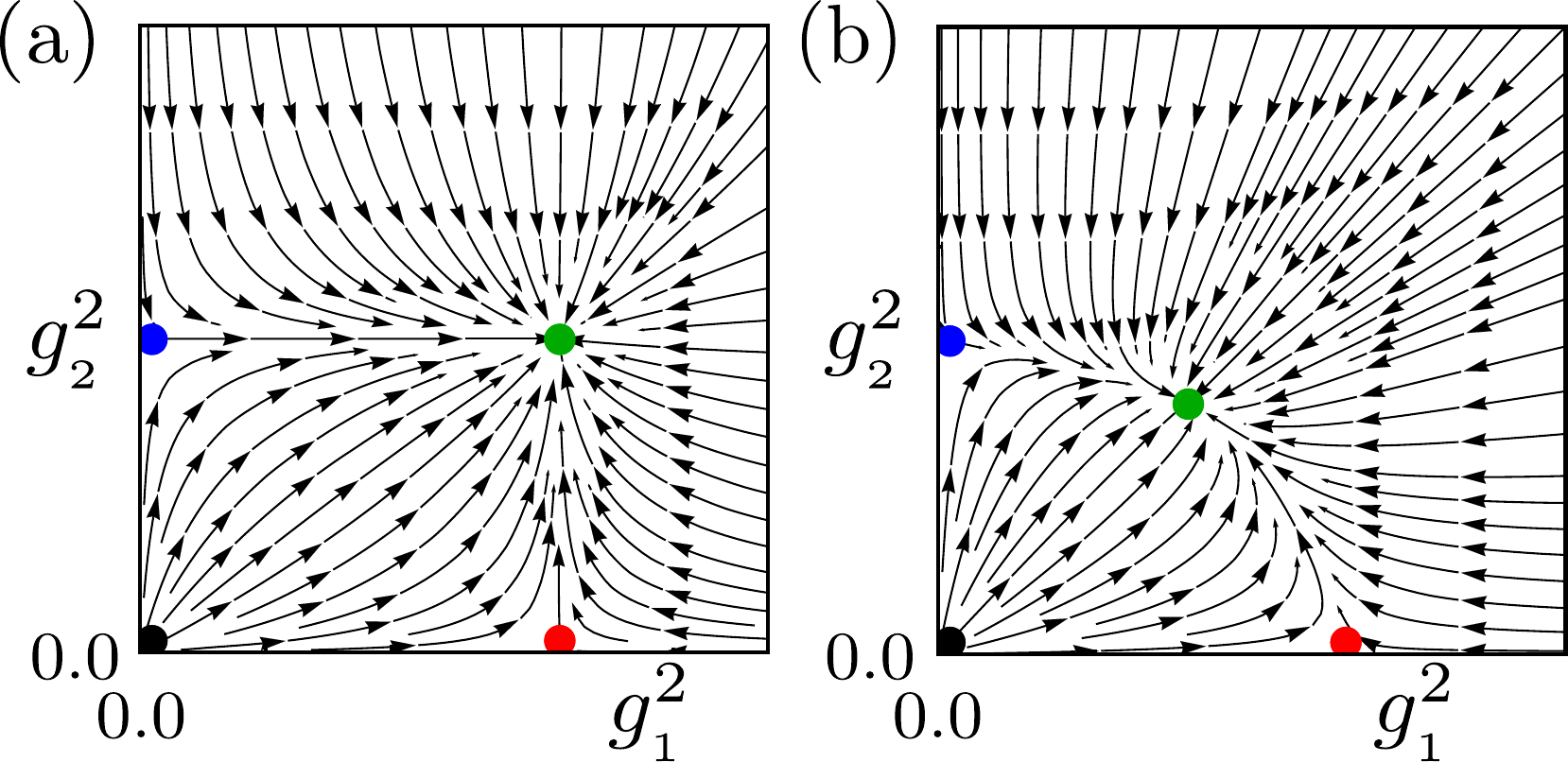}
\caption{{\bf Schematic renormalization-group (RG) flow in the Yukawa plane.} The flow for (a) non-Hermitian and (b) Hermitian Dirac systems [Eqs.~\eqref{eq:yukawaflow1} and~\eqref{eq:yukawaflow2}]. The Yukawa couplings $g^2_{_1}$ and $g^2_{_2}$ are always measured in units of the RG parameter $\varepsilon$. Colored dots correspond to RG fixed points. The red [blue] fixed points in both (a) and (b)  control continuous quantum phase transition (QPT) from Dirac semimetal (DSM) to commuting class mass (CCM) [anticommuting class mass (ACM)] order, tuned by the bosonic mass $m^2_1$ [$m^2_2$].  The flow equations of the bosonic masses are shown in Eq.~\eqref{eq:massflow}. The green fixed point in (a) with $g^2_{_1} \neq g^2_{_2}$ controls a first order transition between the CCM and ACM, which can be tuned by one of the two bosonic masses, and thus devoid of any single-parameter scaling. Their distinct RG flow equations are shown in Eq.~\eqref{eq:massflow}. By contrast, green dot in (b) with $g^2_{_1}= g^2_{_2}$ and an enlarged $O(n)$ symmetry drives a continuous DSM-mixed-class-mass  QPT, tunable by a single mass parameter $m^2=m^2_1=m^2_2$ with its RG flow equation shown in Eq.~\eqref{eq:massflowMCP}.
}~\label{fig:streamplot}
\end{figure}

In this work, we establish a general framework for constructing symmetry-protected Lorentz-invariant NH DSMs  across various dimensions, described by the Dirac Hamiltonianlike operator,  Eq.~\eqref{eq:DiracNH}, which  is achieved by introducing a masslike anti-Hermitian Dirac operator alongside its Hermitian counterpart. Consequently, the thermodynamic, transport, and elastic responses of such NH DSMs closely resemble those of conventional Dirac materials, but in terms of an effective Fermi velocity [Eq.~\eqref{eq:DiracNH}]. 
By employing a leading order $\epsilon=3-d$ expansion about $d=3$ upper critical dimension of the Gross-Neveu-Yukawa quantum-critical  theory, we show that the system  in the vicinity of a NH DSM-insulator quantum phase transition  features an  emergent  Yukawa-Lorentz symmetry in terms of a unique velocity,  as explicitly shown in Eqs.~\eqref{eq:velocity-fermi}-\eqref{eq:flow-anticomm-2}. See also Figs.~\ref{fig:Self-energy} and \ref{fig:TerminalVelocity}. Finally, we unfold the quantum (multi-)critical phenomena in correlated NH DSMs (Fig.~\ref{fig:streamplot}), and characterize it through the critical exponents, such as  bosonic and fermionic anomalous dimensions [Eq.~\eqref{eq:anomalousdim}], and correlation length exponent [Eq.~\eqref{eq:correlationexpo}].

\noindent
{\bf Results and Discussion}

\noindent
{\bf Minimal model}

\noindent 
We set out to construct a minimal effective Hamiltonianlike NH Dirac operator ($H_{\rm NH}$), describing a collection of linearly dispersing gapless quasiparticles in $d$ spatial dimensions coupled to environment, such that $H_{\rm NH}$ possesses either purely real or purely imaginary eigenvalues. We begin with the standard Dirac Hamiltonian of the form $H=\sum_{{\bf k}} \Psi^\dagger_{\bf k} H_{\rm D} \Psi_{\bf k}$, where
\begin{equation}\label{eq:DiracHerm}
H_{\rm D}=v_{_{\rm H}} \; \sum_{j=1}^d \Gamma_j k_j \equiv v_{_{\rm H}} h_0,
\end{equation}
$v_{_{\rm H}}$ bears the dimension of the Fermi velocity, Hermitian $\Gamma$ matrices satisfy the anticommuting Clifford algebra $\{ \Gamma_j, \Gamma_k \}=2 \delta_{jk}$ for $j,k=1, \cdots, d$, and ${\bf k}$ is momentum, yielding a linear energy-momentum relation $E_{\rm H} ({\bf k})= \pm v_{_{\rm H}} |{\bf k}|$. Internal structure of the Dirac spinor $\Psi_{\bf k}$ depends on the microscopic details of the system. For a minimal four-component Dirac spinor, the maximal number of mutually anticommuting Hermitian matrices is five, out of which three (two) can be chosen to be purely real (imaginary). Although our construction is applicable in arbitrary dimensions, here we primarily concentrate on $d=2$. Then, without any loss of generality, we choose $\Gamma_1$ and $\Gamma_2$ to be purely imaginary.

Possible mass terms, $\Psi^\dagger_{\bf k} M \Psi_{\bf k}$, producing isotropically gapped ordered ground states, are then represented by the Hermitian matrices ($M$) that anticommute with the Dirac Hamiltonian in Eq.~\eqref{eq:DiracHerm}, namely $\{ M,h_0 \}=0$ with $M^2=1$. In $d=2$, there are four such mass matrices for a four-component Dirac system $M \in \{M_1, M_2,M_3, \Gamma_{12}\}$, where $\Gamma_{jk}=i\Gamma_j \Gamma_k$. While $M_j$s are purely real for $j=1,2,3$, each of which breaks the SU(2) chiral symmetry of $H_{\rm D}$, generated by $\{M_{12}, M_{23}, M_{31} \}$, with $M_{jk}=i M_j M_k$, the purely imaginary mass matrix $\Gamma_{12}$ transforms as a scalar under the chiral rotation. It breaks the time-reversal symmetry under ${\mathcal K}$ (complex conjugation).

The crucial observation is that the product of a mass matrix and the Hamiltonian $h_0$ is anti-Hermitian, namely $(Mh_0)^\dagger=-M h_0$. Therefore, we define the NH Dirac operator as a minimal extension of $H_{\rm D}$ [Eq.~\eqref{eq:DiracHerm}] by including such a masslike anti-Hermitian term, leading to
\begin{equation}~\label{eq:DiracNH}
H_{\rm NH}=(v_{_{\rm H}}+v_{_{\rm NH}} M)h_0.
\end{equation}
The real parameter $v_{_{\rm NH}}$, also bearing the dimension of the Fermi velocity, quantifies the strength of the system-to-environment coupling. The spectrum of the NH Dirac operator $E_{\rm NH}({\bf k})=\pm \sqrt{v_{_{\rm H}}^2-v_{_{\rm NH}}^2} |{\bf k}| \equiv \pm v_{_{\rm F}} |{\bf k}|$ is purely real (imaginary) for $v_{_{\rm H}}>v_{_{\rm NH}}$ ($v_{_{\rm H}}<v_{_{\rm NH}}$), where $v_{_{\rm F}}=\sqrt{v_{_{\rm H}}^2-v_{_{\rm NH}}^2}$ is the effective Fermi velocity of NH Dirac fermions. Hereafter, we consider the case with the real energy eigenvalues, unless otherwise stated.

The form of the NH Dirac operator ($H_{\rm NH}$) is restricted by four nonspatial discrete unitary and antiunitary symmetries, among which time-reversal, particle-hole and pseudo-Hermitianity~\cite{Bernard2002}, with the corresponding representations given in Table~\ref{tab:symmetry} for a specific choice of $M=M_1$. As explicitly shown, once the form of $H_{\rm NH}$ is fixed (for a given $M$), none of the four constant mass terms is invariant under all of these symmetries, and therefore cannot be added to $H_{\rm NH}$ without breaking at least one of them. Hence, the nonspatial symmetries protect NH gapless Dirac fermions, irrespective of the specific choice of $M$ and the dimensionality of the system ($d$). See Methods and Supplementary Note 1 for details.

\begin{table*}
\begin{tblr}{width=0.8\linewidth,colspec={|Q[c,m]|[2pt,white]Q[c,m]|[2pt,white]Q[c,m]|[2pt,white]Q[c,m]|[2pt,white]Q[c,m]|[2pt,white]Q[c,m]|[2pt,white]Q[c,m]|[2pt,white]Q[c,m]|[2pt,white]Q[c,m]|[2pt,white]Q[c,m]|[2pt,white]Q[c,m]|[2pt,white]Q[c,m]|},cell{even}{1-12}={white},row{1-3}={white},}
\hline
\SetCell[r=3]{c} Mass terms &  \SetCell[c=11]{c} Symmetries\\
\hline
&\SetCell[c=2]{c}{$\mathcal{T_+}$} 
&&\SetCell[c=2]{c}{$\mathcal{C_+}$}&&{$\mathcal{T_-}$}&\SetCell[c=2]{c}{$\mathcal{C_-}$}&&{$\text{PH}_1$}&{$\text{PH}_2$}&{$\text{PSH}_1$}&{$\text{PSH}_2$}\\
\hline
& ${\mathcal K}$     & $i M_2 M_3 \;{\mathcal K}$          &    $i M_1 M_2\;{\mathcal K}$        &    $i M_1 M_3 \;{\mathcal K}$       &   $M_1\;{\mathcal K}$    &    $M_2\;{\mathcal K}$          &    $M_3\;{\mathcal K}$          &  $M_{1}$         &    $i \Gamma_1 \Gamma_2$       &  $i M_1 M_2$  & $i M_1 M_3$   \\
\hline
$M_1$ &          $\checkmark$     & $\checkmark$           &   $\times$        &  $\times$         &   $\times$    &  $\checkmark$          &    $\checkmark$       &     $\times$      &   $\times$        &  $\times$         &   $\times$        \\
\hline 
$M_2$ &    $\checkmark$       &   $\times$         & $\times$           &  $\checkmark$         &  $\checkmark$     &  $\times$         &   $\checkmark$         &    $\checkmark$        &     $\times$      &    $\times$       &   $\checkmark$        \\
\hline
$M_3$ &    $\checkmark$        &   $\times$        &  $\checkmark$           &   $\times$        &  $\checkmark$      &  $\checkmark$          &      $\times$     &  $\checkmark$          &   $\times$        &  $\checkmark$         &  $\times$          \\
\hline
$i \Gamma_1 \Gamma_2 $ &    $\times$       & $\times$          & $\checkmark$          & $\checkmark$          & $\checkmark$      & $\times$          &  $\times$         & $\times$          &   $\times$        & $\checkmark$           &   $\checkmark$ \\
\hline
\end{tblr}
\caption{{\bf Nonspatial symmetries of the non-Hermitian (NH) Dirac operator.} We here show the symmetries of the NH operator $H_{\rm NH}$ in $d=2$ spatial dimensions with the anti-Hermitian term containing the matrix  $M=M_1$ [Eq.~\eqref{eq:DiracNH}].  $\mathcal{T}_+$ and $\mathcal{C}_+$ ($\mathcal{T}_-$ and $\mathcal{C}_-$) represent time-reversal (antiunitary particle-hole) symmetry, while ${\rm PH}_{1,2}$ (${\rm PSH}_{1,2}$) are unitary particle-hole (pseudo-Hermiticity) symmetry, under which ${\mathcal T}_\pm H_{\rm NH} {\mathcal T}^{-1}_\pm=\pm H_{\rm NH}$, ${\mathcal C}_\pm H^\dagger_{\rm NH} {\mathcal C}^{-1}_\pm=\pm H_{\rm NH}$, ${\rm PH}_{1,2} H_{\rm NH} {\rm PH}^{-1}_{1,2}=-H_{\rm NH}$ and ${\rm PSH}_{1,2} H^\dagger_{\rm NH} {\rm PSH}^{-1}_{1,2}=H_{\rm NH}$~\cite{Bernard2002}. We work with a representation in which Hermitian matrices $M_1$, $M_2$ and $M_3$ ($\Gamma_1$ and $\Gamma_2$) are purely real (imaginary). The symbol $\checkmark$ ($\times$) indicates whether a mass term is even (odd) under a specific symmetry. Notice that none of the four possible uniform mass terms is invariant under all the nonspatial symmetries of $H_{\rm NH}$, and thus can not be added to $H_{\rm NH}$ without breaking at least one of them. A similar symmetry protection for NH Dirac nodes in $d=2$ exists for $M=M_2, M_3$ and $i \Gamma_1 \Gamma_2$ (purely imaginary), and also in three dimensions. See Methods and Supplementary Note 1 for details.
}~\label{tab:symmetry}
\end{table*}

\noindent 
{\bf Scaling}

\noindent 
The form of $H_{_{\rm NH}}$ is manifestly Lorentz invariant, with the dynamical exponent $z=1$, measuring the relative scaling between the energy ($E_{\rm NH}$) and momentum (${\bf k}$). This, in turn, implies that the density of states vanishes in a power-law fashion $\rho(E)\sim|E|^{d-1}/v^d_{_{\rm F}}$. As a consequence, a weak local (Hubbardlike) short-range interaction is irrelevant for $d>1$ and the concomitant quantum phase transition (QPT) into an ordered phase takes place through a strongly coupled quantum critical point (QCP) with its critical behavior described by a Gross-Neveu-Yukawa (GNY) theory, about which more in a moment. Nonetheless, a reduced effective Fermi velocity $v_{_{\rm F}}< v_{_{\rm H}}$ is expected to enhance the mass ordering propensity in NH DSMs in comparison to its counterpart in Hermitian Dirac materials, which we show shortly.

To gain a further insight into the nature of such an NH DSM, next we analyze its response to an external electromagnetic field in terms of the  linear longitudinal optical conductivity at zero temperature and finite frequency ($\omega$) within the Kubo formalism. The component of the current operator in the $l^{\rm th}$ spatial direction is $j_l=(v_{_{\rm H}} + v_{_{\rm NH}} M) \Gamma_l$. The requisite polarization bubble reads $\Pi^{(d)}(i \omega)=\sum^{d}_{j=1}[\Pi^{(d)}_{jj}(i\omega)-\Pi^{(d)}_{jj}(0)]/d$, where
 \begin{equation}
 \Pi^{(d)}_{lm}(i\omega)=- {\rm Tr} \int\frac{d\nu d^d{\bf k}}{(2\pi)^{d+1}} \left[j_l G_{\rm F}(i\omega_+,{\bf k})j_m G_{\rm F}(i\omega,{\bf k})\right],
 \end{equation}
$\omega_+=\omega+\nu$, and $G_{\rm F} (i\omega,{\bf k})=(i \omega + H_{\rm NH})/(\omega^2 + v_{_{\rm F}}^2 k^2)$ is the fermionic Green's function. After the analytic continuation $i\omega\to \omega+i\delta$ with $\delta>0$ to real frequency $\omega$, and using the Kubo formula, we obtain the optical conductivity
\begin{equation}~\label{eq:OC}
\sigma_{lm}^{(2)}(\omega)= N_f \frac{e^2}{h} \frac{\pi}{4}\delta_{lm}
\:\; \text{and} \:\;
\sigma_{lm}^{(3)}(\omega)= N_f \frac{e^2}{h} \frac{\omega}{6 v_{_{\rm F}}}\delta_{lm},
\end{equation}
in $d=2$ and $d=3$, respectively. Here $N_f$ is the number of four-component fermion flavors. See Methods and  Supplementary Note 2. Therefore, NH DSMs and its Hermitian counterparts share the same value of the optical conductivity~\cite{LudwigOC1994, GoswamiOC2011, RoyJuricicOC2017}, with the Fermi velocity in $d=3$ replaced by the effective one for NH Dirac fermions ($v_{_{\rm F}}$).

Furthermore, the frequency-dependent optical shear viscosity of noninteracting NH DSMs in $d$ dimensions features a single independent component due to the underlying rotational symmetry, and is given by $\eta^{(d)}_{ijkl}(\omega)={\mathcal P}_{ijkl} \; \eta^{(d)}(\omega)$, where ${\mathcal P}_{ijkl}=\delta_{ik}\delta_{jl}+\delta_{il}\delta_{jk}-(2/d)\delta_{ij}\delta_{kl}$,  
\begin{equation}~\label{eq:viscosity}
\eta^{(2)}(\omega)=\frac{N_f}{128} \left( \frac{\omega}{v_{_{\rm F}}}\right)^2 \: 
\text{and} \;
\eta^{(3)}(\omega)=\frac{N_f}{320\pi} \left( \frac{\omega}{v_{_{\rm F}}}\right)^3. 
\end{equation} 
Therefore, NH and Hermitian DSMs possess the same value of the optical shear viscosity~\cite{MooreViscosity2020}, with the Fermi velocity being replaced by an effective one in the former systems. See Methods and  Supplementary Note 3. These outcomes may imply that the associated conformal field theories feature the same central charge.


\noindent 
{\bf Mass orders}

\noindent
The order parameter (OP) of a uniform Dirac mass that isotropically gaps out Dirac fermions can be expressed as an $n$-component vector $\Phi_j=\langle \Psi^\dagger_{\bf k} N_j \Psi_{\bf k} \rangle$, such that $\{H_{\rm D}, N_j \}=0$ and $\{ N_j, N_k \}=2 \delta_{jk}$ for $j,k= 1, \cdots,n$. But, the appearance of a mass matrix ($M$) in $H_{\rm NH}$ fragments the mass OPs and they can be classified according to the canonical (anti)commutation relations between $N_j$s and $M$: (a) commuting class mass (CCM) for which $[N_j,M]=0$ for all $j$, (b) anticommuting class mass (ACM) with $\{N_j,M\}=0$ for all $j$, and (c) mixed class mass (MCM) for which $[N_j,M]=0$ for $j=1, \cdots, n_1$ and $\{N_j,M\}=0$ for $j=1,\cdots, n_2$ with $n_1+n_2=n$. Hereafter, we mainly focus on the former two classes of mass ordering.

The ordering tendencies toward the nucleation of these two classes of mass ordering, revealing the impact of the NH parameter ($v_{_{\rm NH}}$), can be estimated from their corresponding (nonuniversal) bare mean-field susceptibilities at zero external frequency and momentum, given by
\begin{equation}~\label{eq:susceptibilitiesbare}
\chi_{_1}=N_f\,f(d)\,\frac{\Lambda^{d-1}}{v_{_{\rm F}}} \:\:
\text{and} \:\:
\chi_{_2}=\chi_{_1} \left( 1 + \frac{v^2_{_{\rm NH}}}{v^2_{_{\rm F}}} \right),
\end{equation}
 respectively, where $f(d)=2S_d/[(d-1)(2\pi)^d]$, $S_d=2\pi^{d/2}/\Gamma(d/2)$, and $\Lambda$ is the ultraviolet momentum cutoff up to which the energy-momentum relation remain linear. See Supplementary Note 4. As the effective Fermi velocity in an NH DSM ($v_{_{\rm F}}$) is smaller than its counterpart in Hermitian Dirac materials ($v_{_{\rm H}}$), the bare mean-field susceptibilities are larger in the former system. Given that the (nonuniversal) critical coupling constant ($u^\star$) for any mass ordering is inversely proportional to the bare mean-field susceptibility, NH DSMs are conducive to mass formation at weaker interactions, stemming from its enhanced density of states. However, the competition between CCMs and ACMs requires notion about their associated quantum critical behaviors, captured from the renormalization group (RG) flows of the velocity parameters, $v_{_{\rm H}}$, $v_{_{\rm NH}}$ and $v_{_{\rm F}}$, which we discuss next. This analysis will also shed light on the emergent multi-critical behavior near the MCM condensation. Notice that the dimensionless susceptibility ($\chi v_{_{\rm F}}/\Lambda^{d-1}$) and critical interaction strength ($u^\star /[v_{_{F}} \Lambda^{d-1}]$) are cut-off independent and universal. In fact, the later one can be compared with the critical interaction to the band width ratio in a microscopic setup, which for the N\'eel antiferromagnet order in Hermitian honeycomb lattice is $U^\star/t \approx 4.5 \pm 0.5$~\cite{SorellaHoneycomb1992}, where $t$ ($U$) is the nearest-neighbor hopping amplitude setting the Fermi velocity (on-site Hubbard repulsion).

\noindent 
{\bf Quantum critical theory}

\noindent 
The dynamics of the bosonic OP fluctuations in a GNY quantum critical theory, describing a strong-coupling instability of NH DSMs, is captured by the $\Phi^4$ theory, with the bosonic correlator given by $G_{\rm B}(i\omega,{\bf k})=[\omega^2+v_{_{\rm B}}^2 k^2]^{-1}$, where $v_{_{\rm B}}$ is the velocity of the OP fluctuations. The dynamics of fermions is governed by the NH Dirac operator in Eq.~\eqref{eq:DiracNH}. These two degrees of freedom are coupled through a Yukawa vertex, entering the imaginary time ($\tau$) action as
\begin{eqnarray}\label{eq:yukawa}
\hspace{-0.2cm}
S_Y=g \int d\tau \int d^d{{\bf x}} \sum^{n}_{j=1} \; \Phi_j(\tau,{\bf x})\,\,\Psi^\dagger(\tau,{\bf x}) N_j \Psi(\tau,{\bf x}).
\end{eqnarray}
At the upper critical three spatial dimensions, both Yukawa ($g$) and the $\Phi^4$ coupling ($\lambda$) are marginal~\cite{ZinnJustin2002}. We, therefore, use the distance from the upper critical dimension $\varepsilon=3-d$ as the expansion parameter to capture the low-energy phenomena close to the GNY QCP. The ultraviolet divergences of the Feynman diagrams are captured using the dimensional regularization and the cut-off independent universal physical quantities, such as the RG flow equations and critical exponents, are computed using the method of minimal subtraction~\cite{Peskin2019, ZinnJustin2002}. In particular, we are interested in the fate of an emergent Yukawa-Lorentz symmetry close to such a RG fixed point, since fermionic and bosonic velocities are generically different at the lattice (ultraviolet) scale.

\noindent
{\bf Yukawa-Lorentz symmetry}

\noindent
To this end, we compute the fermionic [Fig.~\ref{fig:Self-energy}(a)] and bosonic [Fig.~\ref{fig:Self-energy}(b)] self-energy diagrams to the leading order in the $\varepsilon$ expansion. Irrespective of the nature of the mass ordering (CCM or ACM), the RG flow of the Hermitian component of the Fermi velocity ($v_{_{\rm H}}$) takes the form
\begin{equation}~\label{eq:velocity-fermi}
\beta_{v_{_{\rm H}}}=-\frac{4 g^2 n}{3 v_{_{\rm B}}(v_{_{\rm F}} + v_{_{\rm B}})^2}\left(1-\frac{v_{_{\rm B}}}{v_{_{\rm F}}}\right)v_{_{\rm H}},
\end{equation}
after rescaling the Yukawa coupling according to $g^2\to {g^2}/{(8\pi^2)}$, where $\beta_Q\equiv dQ/d\ln b$ and $b$ is the RG time. The flow equations for the remaining two velocities, when the mass matrix commutes with $M$ (CCM) are
\begin{align}~\label{eq:flow-comm-1}
\beta_{v_{_{\rm NH}}}&=-\frac{4 g^2 n}{3 v_{_{\rm B}} (v_{_{\rm F}} + v_{_{\rm B}})^2} \left(1-\frac{v_{_{\rm B}}}{v_{_{\rm F}}}\right)v_{_{\rm NH}} \\
\label{eq:flow-comm-2}
\text{and} \: \: \beta_{v_{_{\rm B}}}&=-N_f \; \frac{g^2 n}{2v_{_{\rm F}}^3}\;\left(1-\frac{v_{_{\rm F}}^2}{v_{_{\rm B}}^2}\right)v_{_{\rm B}}.
\end{align}
These RG flow equations [Eq.~\eqref{eq:velocity-fermi}-\eqref{eq:flow-comm-2}] imply that at the GNY QCP with $g^2 \sim \varepsilon$, the terminal velocities are such that  $v_{_{\rm F}}=\sqrt{v_{_{\rm H}}^2-v_{_{\rm NH}}^2}=v_{_{\rm B}}$, with all three velocities being nonzero, independent of their initial values. See Supplementary Note 5. We also confirm this outcome by numerically solving these flow equations. See Fig.~\ref{fig:TerminalVelocity}(a). Therefore, a new fixed point with an enlarged symmetry emerges, at which the system remains coupled to the environment and the effective Fermi velocity of NH Dirac fermions ($v_{_{\rm F}}$) is equal to the bosonic OP velocity ($v_{_{\rm B}}$), with the nonzero terminal values for $v_{_{\rm H}}$, $v_{_{\rm NH}}$ and $v_{_{\rm B}}$. We name it non-Hermitian Yukawa-Lorentz symmetry.

On the other hand, when a NH DSM arrives at the brink of the ACM condensation, the flow equations for $v_{_{\rm NH}}$ and $v_{_{\rm B}}$ take the forms (see Supplementary Note 5)
\begin{align}\label{eq:flow-anticomm-1}
\beta_{v_{_{\rm NH}}}&=-\frac{8 g^2 n}{3 v_{_{\rm B}} (v_{_{\rm F}} + v_{_{\rm B}})^2} \left(1 + \frac{v_{_{\rm B}}}{2 v_{_{\rm F}}} \right) v_{_{\rm NH}}, \\
\label{eq:flow-anticomm-2}
\text{and} \: \: \beta_{v_{_{\rm B}}} &=-N_f \; \frac{g^2 n}{2 v_{_{\rm F}}^3} \left(\frac{v_{_{\rm H}}^2}{v_{_{\rm F}}^2}-\frac{v_{_{\rm H}}^2}{v_{_{\rm B}}^2}-\frac{2 v_{_{\rm NH}}^2}{3 v_{_{\rm B}}^2}\right) v_{_{\rm B}},
\end{align}
respectively. These two flow equations along with Eq.~\eqref{eq:velocity-fermi} imply that near the ACM class GNY QCP ($g^2 \sim \varepsilon$), the system recovers Hermiticity by decoupling itself from the environment and a conventional Yukawa-Lorentz symmetry emerges with $v_{_{\rm NH}}=0$ and $v_{_{\rm F}}=v_{_{\rm H}}=v_{_{\rm B}}$, irrespective of their bare values. We further confirm this outcome by numerically solving the flow equations. See Fig.~\ref{fig:TerminalVelocity}(b).

\noindent 
{\bf Mean-field theory}

\noindent 
Guided by the emergent quantum critical theory, we now compare the tendencies between CCM and ACM orderings. At the corresponding GNY QCPs, their renormalized susceptibilities are respectively
\begin{equation}~\label{eq:susceptibilitiesrenor}
\chi^\star_{1}=N_f\,f(d)\,\frac{\Lambda^{d-1}}{v_{_{\rm F}}} \:\:
\text{and} \:\:
\chi^\star_{2}=N_f\,f(d)\,\frac{\Lambda^{d-1}}{v_{_{\rm H}}},
\end{equation}
obtained from Eq.~\eqref{eq:susceptibilitiesbare} by replacing various velocities with their terminal ones. As $v_{_{\rm F}}<v_{_{\rm H}}$ in NH DSMs, $\chi^\star_{_1}>\chi^\star_{_2}$, and thus $u^\star_{_1}<u^\star_{_2}$, where $u^\star_{_1}$ ($u^\star_{_2}$) is the critical strength of the coupling constant for the CCM (ACM) ordering. This outcome is particularly important when the mass matrix $M$, appearing in $H_{\rm NH}$ [Eq.~\eqref{eq:DiracNH}], is a component of a vector mass order of a Hermitian Dirac system. Any such choice of $M$ immediately fragments the vector mass order into CCM and ACM, and our susceptibility analysis suggests that the OP corresponding to the CCM enjoys a stronger propensity toward nucleation.

This prediction can be tested from the quasiparticle spectra of NH Dirac fermions inside mass ordered phases, described by an effective NH single-particle operator
\begin{equation}
H_{\rm NH}^{\rm MF}=(v_{_{\rm H}}+v_{_{\rm NH}} M)h_0 + \sum_{j} \Delta_j N_j,
 \end{equation}
where $N_j$s are the Hermitian mass matrices, satisfying $\{ N_j,h_0 \}=0$ for all $j$, and $\Delta_j$ is the OP amplitude. The excitation spectrum of $H_{\rm NH}^{\rm MF}$ crucially depends on the nature of the mass ordering. Namely, (a) for $[N_j,M]=0$ (CCM), the energy eigenvalues are either purely real ($v_{_{\rm H}}>v_{_{\rm NH}}$) or complex ($v_{_{\rm H}}<v_{_{\rm NH}}$), whereas (b) for $\{N_j,M\}=0$ (ACM) the excitation spectrum is complex for any $v_{_{\rm H}}$ and $v_{_{\rm NH}}$. Next we exemplify these outcomes by focusing on a specific microscopic scenario.

\noindent
{\bf NH honeycomb Hubbard model}

\noindent
Consider a collection of massless Dirac fermions on graphene's honeycomb lattice. In this system, the Ne\'el antiferromagnetic (AFM) order corresponds to the three-component vector mass that breaks the O(3) spin rotational symmetry, favored by a strong on-site Hubbard repulsion at half-filling~\cite{SorellaHoneycomb1992} due to the underlying bipartite nature of the lattice allowing the formation of staggered magnetization without any frustration. We construct an NH Dirac operator by choosing the easy $z$-axis component of the AFM order as $M$, which then naturally becomes a CCM order. By contrast, two planar or easy-plane components of the AFM order fall within the category of ACM. See Supplementary Note 6. Our susceptibility calculation predicts that such an NH Hubbard model should prefer nucleation of an Ising symmetry breaking easy-axis AFM order over the XY symmetry breaking easy-plane one, and the quasi-particle spectra inside the ordered phase must be purely real unless $v_{_{\rm H}}<v_{_{\rm NH}}$ at the bare level. These predictions can serve as the litmus tests for our quantum critical theory of correlated NH Dirac materials in quantum Monte Carlo simulations and exact numerical diagonalizations~\cite{NHHubbard2023-1}.

\noindent
{\bf NH criticality}

\noindent
Finally, we compute the critical exponents near the GNY QCPs associated with the $n_1$-component CCM and $n_2$-component ACM orderings. The RG flow equations for the corresponding Yukawa couplings respectively read as
\begin{eqnarray}
\label{eq:yukawaflow1}
\beta_{g^2_{_1}} &=& \varepsilon g^2_{_1} - (2 N_f +4 -n_1) g^4_{_1} + n_2 g^2_{_1} g^2_{_2} \delta_{v^0_{_{\rm NH}},0},\\
\label{eq:yukawaflow2}
\beta_{g^2_{_2}} &=& \varepsilon g^2_{_2}- (2 N_f +4 -n_2) g^4_{_2} + n_1 g^2_{_1} g^2_{_2} \delta_{v^0_{_{\rm NH}},0},
\end{eqnarray}
after rescaling the coupling constants according to $g^2_{_1}/(8 \pi^2 v^2_{_{\rm F}}) \to g^2_{_1}$ and $g^2_{_2}/(8 \pi^2 v^2_{_{\rm H}}) \to g^2_{_2}$. See Supplementary Note 7 for details. The Yukawa fixed points at $g^2_{_{1,\star}}=\varepsilon/[2 N_f+4-n_1]$ [red dot in Fig.~\ref{fig:streamplot}(a)] and $g^2_{_{2,\star}}=\varepsilon/[2 N_f+4-n_2]$ [blue dot in Fig.~\ref{fig:streamplot}(a)] respectively control the continuous QPTs into the CCM and ACM orders. At the Yukawa fixed points the fermionic and bosonic anomalous dimensions are respectively
\begin{equation}~\label{eq:anomalousdim}  
\eta_{_{\Psi,j}}= \frac{n_j}{2} g^2_{_{j,\star}}
\:\: \text{and} \:\:
\eta_{_{\Phi,j}} = 2 N_f g^2_{_{j,\star}}
\end{equation}
for $j=1,2$, and the fermionic Green's functions scale as $G^{-1}_{\rm F}(b) \sim (\omega^2 +v^2_{_{\rm F}}(b) |{\bf k}|^2)^{1-\eta_{_{\Psi,1}}}$ and $(\omega^2 +v^2_{_{\rm H}}(b) |{\bf k}|^2)^{1-\eta_{_{\Psi,2}}}$, indicating the onset of non-Fermi liquids therein. The bosonic Green's function scale as $G^{-1}_{\rm B}(b) \sim (\omega^2 +v^2_{_{\rm B}}(b) |{\bf k}|^2)^{1-\eta_{_{\Phi,j}}}$. While the universal critical exponents are independent of the velocity parameters, their RG flows shown in Fig.~\ref{fig:TerminalVelocity} determine the scale-dependent ($b$) fermionic and bosonic Green's or spectral function within the quantum critical regime. Here $b$ can be identified as $b=\Lambda/\Lambda_0$, where $\Lambda(\Lambda_0)$ is the running (fixed lattice or ultraviolet) momentum or energy scale.

The last terms in Eqs.~\eqref{eq:yukawaflow1} and~\eqref{eq:yukawaflow2} are pertinent in the proximity to an $n$-component MCM ordering, where $n=n_1 + n_2$, and they are nontrivial only when the bare value of the NH Fermi velocity ($v^0_{_{\rm NH}}$) is zero, corresponding to a Hermitian Dirac system. Then a fully stable quantum multi-critical point with enlarged $O(n)$ symmetry emerges at $g^2_{_{1,\star}}=g^2_{_{2,\star}}=\varepsilon/[2 N_f+4-n]$~\cite{BRoy2011, RoyGoswamiJuricic2018}. These terms, however, do not exist in NH Dirac system as CCM and ACM components of the MCM order possess distinct Yukawa-Lorentz symmetries (hence cannot be coupled). The fixed point located at $(g^2_{_{1,\star}},g^2_{_{2,\star}})=(\varepsilon/[2 N_f+4-n_1],\varepsilon/[2 N_f+4-n_2])$ thus controls a direct first-order phase transition between them, as the transition can be tuned by either one of the bosonic masses $m^2_1$ or $m^2_2$, for which the distinct RG flow equations are shown below in Eq.~\eqref{eq:massflow}. We do not discuss such first-order transition between two ordered phases in any further detail. These two scenarios are shown in Fig.~\ref{fig:streamplot}.

The RG flow equations for the $\Phi^4$ couplings associated with the CCM and ACM orderings assume the forms
\begin{equation}\label{eq:flow-lambda}
\beta_{\lambda_j}=\varepsilon \lambda_j - 4 N_f g^2_{_j} \left[ \lambda_j - 6 g^2_{_j} \right] - \frac{8+n_1}{6} \lambda^2_j
\end{equation}
for $j=1$ and $2$, respectively, in terms of the rescaled coupling constant $\lambda_1/(8 \pi^2 v_{_{\rm F}}) \to \lambda_1$ and $\lambda_2/(8 \pi^2 v_{_{\rm H}}) \to \lambda_2$. The fixed point values $\lambda_{j,\star}$ (somewhat lengthy expression) are obtained from the solution of $\beta_{\lambda_j}=0$. The RG flow equation for the bosonic mass parameters that serve as the tuning parameter for the NH DSM to CCM and ACM QPTs respectively take the forms
\begin{equation}~\label{eq:massflow}
\beta_{m^2_j}= {m^2_j} \left[2- 2 N_f g^2_{_{j}} - \frac{2+n_j}{6} \lambda_j \right]
\end{equation}
for $j=1$ and $2$, yielding the correlation length exponents
\begin{equation}~\label{eq:correlationexpo}
\nu_j=\frac{1}{2} + \frac{N_f}{2} g^2_{_{j,\star}} + \frac{2+n_j}{24} \lambda_{j,\star}.
\end{equation}
On the other hand, the RG flow equation for the mass parameter controlling the continuous QPT between two ordered phases across the multi-critical point [green dot in Fig.~\ref{fig:streamplot}(b)] in Hermitian Dirac system reads as 
\begin{equation}~\label{eq:massflowMCP}
\beta_{m^2}={m^2} \left[2- 2 N_f g^2 - \frac{2+n}{6} \lambda \right],
\end{equation}
where $n=n_1+n_2$, $g^2=g^2_{_1}=g^2_{_2}$, $\lambda=\lambda_1=\lambda_2$ and $m^2=m^2_1=m^2_2$, reflecting the emergent $O(n)$ symmetry.

\noindent\\
 {\bf Conclusions}

\noindent
In this work, we develop a general formalism of constructing symmetry protected Lorentz invariant NH DSMs in any dimension by introducing masslike anti-Hermitian Dirac operator to its Hermitian counterpart, featuring purely real or imaginary eigenvalue spectra. Their thermodynamic, transport and elastic responses closely mimic the ones in conventional Dirac materials, however in terms of an effective Fermi velocity $v_{_{\rm F}}=\sqrt{v^2_{_{\rm H}} - v^2_{_{\rm NH}}}$ of NH Dirac fermions, where $v_{_{\rm H}}$ ($v_{_{\rm NH}}$) is the Fermi velocity associated with the Hermitian (anti-Hermitian) component of the NH Dirac operator. Following Eq.~\eqref{eq:DiracNH}, our construction of the NH DSMs can be generalized to any semimetal of arbitrary energy dispersion relation, described by an Hermitian operator ($h_0$) and accommodates mass orders ($M$), to construct its NH counterpart, since in all these cases the anti-Hermitian operator $M h_0$ exists.

We show that when NH DSMs arrive in the close proximity to any mass ordering via spontaneous symmetry breaking, the emergent non-Fermi liquid features NH (for CCM) or Hermitian (for ACM) Yukawa-Lorentz symmetry in terms of a unique terminal velocity for all the participating degrees of freedom. We also determine the non-trivial critical exponents associated with the NH DSM to Dirac insulator QPTs, revealing their non-Gaussian nature in $d=2$, favored by strong Hubbardlike local interactions. Combining mean-field theory and RG analysis, we address the competition among different classes of mass orderings and argue that the nature of the Dirac insulators can be identified from the quasiparticle spectra inside the ordered phases, which can serve as a direct test of our proposed effective description of correlated NH DSMs in numerical simulations~\cite{NHHubbard2023-1} and experiments. In three spatial dimensions, the NH DSM to insulator QPTs are Gaussian in nature as they take place at $g^2_{_j}=\lambda_{_j}=0$ since $\varepsilon=0$ therein, yielding mean-field critical exponents $\nu=1/2$ and $\eta_{_\Psi}=\eta_{_\Phi}=0$. Still, the fermionic and bosonic quasiparticle poles suffer logarithmic corrections, producing a marginal Fermi liquid, which also manifests the emergent Yukawa-Lorentz symmetry.

Even though the universal critical exponents are independent of any velocity parameters and microscopic details of the system, and only depends on the parameter $\varepsilon$, the nature of the order phases (characterized by the number of bosonic order-parameter components $n$), and the four-component Dirac fermionic flavor number $N_f$, the signatures of $v_{_{\rm H}}$, $v_{_{\rm NH}}$ and $v_{_{\rm B}}$ can be probed from the energy dependent measurement of the two-point fermionic ($G_{\rm F}$) and bosonic ($G_{\rm B}$) correlation functions, capturing the RG flow of these parameters shown in Fig.~\ref{fig:TerminalVelocity}. Furthermore, the non-Hermiticity can also change the nature of the ordered phases in NH DSMs in comparison to its counterpart in Hermitian DSMs and that way directly affecting the associated quantum critical phenomena. For example, Hubbard repulsion in Hermitian honeycomb lattice supports O(3) symmetry breaking N\'eel antiferromagnet order, with the critical exponents determined by setting $N_f=2$ and $n=3$. If we choose the mass matrix $M$ to be the $z$ or easy-axis component of the N\'eel order in the construction of the NH Dirac operator $H_{\rm NH}$ [Eq.~\eqref{eq:DiracNH}], the ultimate ordered state is expected to be the Ising symmetry breaking easy-axis antiferromagnet. Therefore, the critical exponent for such an NH honeycomb Hubbard model is set by $N_f=2$ and $n=1$.

For experimental realizations of our proposal, consider a paradigmatic Dirac system in $d=2$, graphene~\cite{Castroneto2009}, where conventional Dirac Hamiltonian ($h_{0}$) results from electronic hopping between the nearest-neighbor sites of $A$ and $B$ sublattices. As such graphene can accommodate a plethora of Dirac masses~\cite{Ryu2009, Szabo2021} and any one of them can be chosen as $M$ in $H_{\rm NH}$ [Eq.~(\ref{eq:DiracNH})]. Consider the simplest one, charge-density-wave with staggered pattern of electronic density between two sublattices~\cite{Semenoff1984}. Then the anti-Hermitian operator $M h_0$ also corresponds to nearest-neighbor hopping on the honeycomb lattice. But, the hopping strength from $A$ to $B$ sites is stronger or weaker than that along the opposite direction, yielding non-Hermiticity. See Supplementary Note 8. Such a simple NH Dirac operator can be engineered in designer electronic and optical honeycomb lattices, on which Hermitian graphene has already been realized~\cite{Manoharan2012, Esslinger2013}. In light of the recent proposal of realizing NH one-dimensional optical lattice with a left-right hopping imbalance~\cite{GongPRX2018}, a hopping imbalance between the $A \to B$ and $B \to A$ directions can be engineered with two copies of the honeycomb lattice, occupied by neutral atoms living in the ground state and first excited state, which are coupled by running waves along three nearest-neighbor bond directions and the sites constituted by the excited state atoms undergoes a rapid loss. When the wavelength of the running wave is equal to the lattice spacing, a NH Dirac operator on optical honeycomb lattice with hopping imbalance between $A \to B$ and $B \to A$ directions can be realized. We point out  that on one-dimensional optical lattices NH models with coupling to the environment has recently been demonstrated~\cite{QianLiang2022}. A similar construction can possibly be engineered on designer electronic lattices to mimic the desired hopping imbalance.

Given that Hubbard repulsion driven AFM ordering has been observed on optical graphene~\cite{Esslinger2013} and the tunable NH coefficient $v_{_{\rm NH}}$ reduces the critical interaction for this ordering, since AFM is a CCM when $M$ is the charge-density-wave order, our predicted mass nucleation at weak coupling and the associated quantum critical phenomena should be observed in these NH designer Dirac materials. Simplicity of this construction should also make the proposed phenomena observable at least on three-dimensional NH optical Dirac lattices, which also supports a number of mass orders~\cite{SzaboJHEP2021}.

Altogether our Lorentz invariant construction of the NH Dirac operator constitutes an ideal theoretical framework to extend the realm of various exotic many-body phenomena in the presence of system-to-environment interactions. Among them quantum electrodynamics, topological defects in the ordered phases, magnetic catalysis, superconductivity, chiral anomaly, to name a few in NH Dirac materials, are the prominent and fascinating ones. The present work lays the foundation of systematic future investigations of these avenues.
\\

\noindent\\
{\bf Methods}

\noindent
In this section, we outline the key methodologies employed in this work to arrive at various conclusions.\\

\noindent
{\bf Symmetry analysis of the NH Dirac operator}

\noindent
The symmetry analysis of the NH Dirac operator, given by Eq.~\eqref{eq:DiracNH} is straightforwardly performed, by using the corresponding canonical (anti)commutation relations of the $16$  Hermitian $4\times4$  Dirac matrices squaring to one, in both $d=2$ and $d=3$ spatial dimensions. For this analysis, it is also of crucial importance to recall that out of these $16$ Dirac matrices maximally five of them mutually anticommute. In this set of five mutually anticommuting matrices, three are purely real while two are purely imaginary. Now, the structure of the purely Hermitian Dirac Hamiltonian depends on the dimensionality. In $d=2$, the two Dirac matrices in the kinetic term can be chosen to be purely imaginary, while in $d=3$ time-reversal symmetry enforces  the three Dirac matrices in the kinetic to be purely real. In either case, one of the remaining Dirac matrices can be chosen to form the anti-Hermitian piece of the NH Dirac operator.  Given the above, we classify the NH Dirac Hamiltonian according to the possible symmetries of the NH operators~\cite{Bernard2002}. This procedure leads to the classification presented in Table~\ref{tab:symmetry} for NH Dirac operator in $d=2$, while the analogous classification for the $d=3$ is given in Supplementary Note 1, together with the additional details, summarized in Supplementary Table~1 and Supplementary Table~2. \\

\noindent
{\bf Zero temperature optical conductivity}

\noindent
The optical conductivity at zero temperature is computed using the standard Kubo formula, which, for the completely isotropic system, considered here, takes the form,  
\begin{equation}\label{Seq:OC-2}
  \sigma(\omega)=\left.\frac{{\rm Im}\;\Pi(i\omega\to\omega+i\eta)}{\omega}\right\vert_{\eta\to0},
\end{equation}
where $\Pi(i\omega)$ is the diagonal element of the polarization tensor of the noninteracting NH Dirac fermions after subtracting its zero-frequency part, $\Pi(i\omega)=(1/d)\sum_{j=1}^{d}[\Pi_{jj}(i\omega)-\Pi_{jj}(0)]$. See Supplementary Figure~1(a). Notice that $\sigma(\omega)$ is computed  at real frequency $\omega$ after performing the analytical continuation from the Matsubara frequency $i\omega\to \omega+i\delta$ in the limit $\delta\to0$.

The polarization tensor explicitly reads as
\begin{align}\label{Meteq:polarization}
\Pi_{jj} (i \omega) &= -{\rm Tr}\int_{-\infty}^\infty\frac{d\nu}{2\pi}\int\frac{d^d{\bf k}}{(2\pi)^d} \left[J_j\;G_{\rm F}(i\omega+i\nu,{\bf k})\right. \nonumber \\
&\qquad \left.\times J_j\; G_{\rm F}(i\nu,{\bf k})\right],
\end{align}
where $j=1, \cdots ,d$,  the current operator $J_j=(v_{_{\rm H}} + v_{_{\rm NH}}M) \Gamma_j$, and
\begin{equation}~\label{Meteq:DiracGF}
G_{\rm F} (i\omega,{\bf k})=\frac{i \omega + H_{\rm NH}}{\omega^2 + v_{_{\rm F}}^2 k^2}
\end{equation}
is the fermionic Green's function corresponding to the NH Dirac operator in Eq.~\eqref{eq:DiracNH} with $v_{_{\rm F}}^2=v_{_{\rm H}}^2-v_{_{\rm NH}}^2$. The rotational symmetry implies isotropy of the polarization tensor. The straightforward calculations then yield the results in Eq.~\eqref{eq:OC} for the zero-temperature optical conductivity  in $d=2$ and $d=3$. Additional details are given in Supplementary Note 2. \\

\noindent
{\bf Zero temperature shear viscosity}

\noindent
In the case of the optical shear viscosity tensor,  we use the corresponding Kubo formula, given by 
\begin{equation}\label{Meteq:eta-Kubo}
\eta^{(d)}_{ijkl}(\omega)=\lim_{\delta\to0}\frac{C^{(d)}_{ijkl}(i\omega\to \omega+i\delta)}{\omega}, 
\end{equation}
in terms of the stress tensor correlation function
\begin{align}\label{Meteq:stress-tensor-Kubo}
C^{(d)}_{ijkl}(i\omega)&=-{\rm Tr}\int_{-\infty}^\infty\frac{d\nu}{2\pi}\int\frac{d^d{\bf k}}{(2\pi)^d}\left[G_{\rm F}(i\nu,{\bf k})\;T_{ij}({\bf k})\right. \nonumber \\
&\qquad \left.\times
G_{\rm F}(i\nu+i\omega,{\bf k})\;T_{kl}({\bf k})\right].
\end{align}
Here,  $T_{lm}({\bf k})=T_{lm}^{({\rm o})}+T_{lm}^{({\rm s})}$ are the orbital and the spin parts of the stress tensor, respectively, given by
\begin{equation}
 T_{lm}^{({\rm o})}({\bf k})=k_m \frac{\partial H_{\rm NH}({\bf k})}{\partial k_l}, \;\;    T_{lm}^{({\rm s})}({\bf k})=i[\mathcal{S}_{lm},H({\bf k})],
\end{equation}
with $\mathcal{S}_{lm}=(i/8)[\Gamma_l,\Gamma_m]$ representing the generators of the spin rotations, and  $i,j,k$ and $l$ are the spatial indices. In particular, in $d=2$, the generator of spin rotations takes the form $S_{12}=(i/4)\Gamma_1\Gamma_2$, while in $d=3$, the three generators can be explicitly written as  $\mathcal{S}_{lm}=(i/4)\Gamma_l\Gamma_m$, with $l\neq m$, and $l,m=1,2,3$. The straightforward calculations then yield Eq.~\eqref{eq:viscosity}. The additional details are given in Supplemental Note 3. \\

\noindent
{\bf RG flow equations in the GNY theory}

\noindent
We first outline the details of the RG analysis leading to the form of the RG flow equations of the fermionic and bosonic  velocities, given by  Eqs.~\eqref{eq:velocity-fermi}-\eqref{eq:flow-anticomm-2}. To this end, we employ the leading-order $\varepsilon$ expansion about $d=3$ upper critical spatial dimensions of the NH Gross-Neveu-Yukawa (GNY) theory with $\varepsilon=3-d$. We compute the self-energy diagrams for the fermions and bosons, shown in~\ref{fig:Self-energy}(a) and (b), respectively, while the Yukawa boson-fermion interaction is given by Eq.~\eqref{eq:yukawa}. See also Supplementary Figures~1(b) and (c). The fermionic self-energy takes the form
\begin{align}
   & \Sigma(i\nu,{\bf k})=g^2\int\frac{d^d{\bf q}}{(2\pi)^d}\int\frac{d\omega}{2\pi} N_i G_{\rm F}(i\omega,{\bf q}) \nonumber \\
& \times
    N_i G_{\rm B} (i(\nu-\omega),{\bf k}-{\bf q}) \equiv g^2 N_i I(i\nu,{\bf k}) N_i,
\end{align}
while the bosonic self-energy is 
\begin{align}
    &\Pi(i\nu,{\bf k})=-\frac{g^2}{2} \int\frac{d^d{\bf q}}{(2\pi)^d}\int\frac{d\omega}{2\pi}\;{\rm Tr} \left[N_i\; G_{\rm F} (i\omega,{\bf q}) \; N_i 
 \right. \nonumber \\
& \left.\times G_{\rm F}(i\nu+i\omega,{\bf k}+{\bf q})\right].
\end{align}

The fermionic Green's function is given by Eq.~\eqref{Meteq:DiracGF}, while the bosonic propagator is
\begin{equation}~\label{Meteq:BosoncGF}
G_{\rm B} (i\omega,{\bf k})=\frac{1}{\omega^2 + v_{_{\rm B}}^2 k^2}.
\end{equation}

We then employ the minimal subtraction scheme~\cite{ZinnJustin2002}, as detailed in Supplementary Note 5, to find the flow equations of the  fermionic and bosonic  velocity parameters, both when the order parameter is CCM and ACM,  ultimately yileding  Eqs.~\eqref{eq:velocity-fermi}-\eqref{eq:flow-anticomm-2}. Additional technical details are given in Supplemental Note 5. 

Analogously, we find the flow equations of the Yukawa  and the $\Phi^4$-vertex, by computing the corresponding diagrams, respectively, shown in Supplementary Figure 2(a) and Supplementary Figures 2(b) and (c). We then employ the minimal subtraction scheme to extract the corresponding renormalization constants, which ultimately yield the RG flow equations in Eqs.~\eqref{eq:yukawaflow1}, \eqref{eq:yukawaflow2}, and \eqref{eq:flow-lambda}. Additional details are shown in Supplemental Note 6. \\

\noindent\\
{\bf Data Availability}\\
All the calculation details are provided in Supplementary Information.

\noindent\\
{\bf Code Availability}\\
Not applicable.

\noindent\\
{\bf Acknowledgments}\\
V.J. acknowledges support of the Swedish Research Council (VR 2019-04735) and Fondecyt (Chile) Grant No. 1230933.  B.R. was supported by NSF CAREER Grant No.\ DMR-2238679. Nordita is partially supported by Nordforsk.

\noindent\\
{\bf Author Contributions}\\
V.J.\ and B.R.\ performed all the calculations and wrote the manuscript. B.R.\ conceived and structured the project.

\noindent\\
{\bf Competing interests}\\
The authors declare no conflicts of interest.

\bibliographystyle{naturemag}

\bibliography{references}

\end{document}